\documentclass[preprint]{revtex4}
\usepackage{graphicx}

\begin{document}

\baselineskip=21.5pt

\begin{flushleft}

{\Large {\bf Desynchronization  and sustainability  of noisy
metapopulation cycles }}

\vskip 1.0cm

{\large Refael Abta$^{1}$, Marcelo Schiffer$^{2}$, Avishag
Ben-Ishay$^{1}$   and Nadav M. Shnerb$^{1}$}

\begin{it}
($1$)  Department of Physics, Bar-Ilan University, Ramat-Gan 52900
Israel  \\ ($2$) Department of Physics, Judea and Samaria College,
Ariel 44837 Israel.
\end{it}

\vskip 5.0cm

$^1$Fax: 972-3-5317630.  E-mail: shnerbn@mail.biu.ac.il.

\vskip 3.0cm A short running head title: Stable metapopulation
cycles. \vskip 1.0cm
 \textbf{Key Words}: coexistence, competition, noise, spatial
 models, predation, diversity, dispersal, desynchronization.
\vskip 1.0cm
Corresponding author: Nadav Shnerb, Department of
Physics, Bar-Ilan University, Ramat-Gan 52900 Israel.

\end{flushleft}

\newpage

\begin{flushleft}
\textbf{Abstract}
\end{flushleft}

The apparent stability of population oscillations in ecological
systems is a long-standing puzzle.  A generic solution for this
problem is suggested here. The stabilizing mechanism involves the
combined effect of spatial migration, amplitude-dependent frequency,
and noise: small differences between spatial patches, induced by the
noise, are amplified by the frequency gradient. Migration among
desynchronized regions then stabilizes the oscillations in the
vicinity of the homogenous manifold. A simple model of diffusively
coupled oscillators allows the derivation of quantitative results,
like the  dependence of the desynchronization upon diffusion
strength and frequency differences.  For an unstable system, a
noise-induced transition is demonstrated, from extinction for small
noise to stability if the noise exceeds some threshold. The coupled
oscillator model is shown to reproduce all previously suggested
stabilizing mechanisms. Accordingly, we suggest using this model as
a standard  tool for identification and classification of population
oscillations on spatial domains.

\newpage

\section{Introduction}

The apparent stability of prey-predator systems is an age-old
puzzle. While natural host-parasitoid and prey-predator systems
persist for millions of years without extinction, simple
considerations imply that this situation should be inherently
unstable.  Essentially, when a predator consumes a prey, it clearly
increases its fitness and its chance to breed and produce more
offspring. Thus, the  predation rate should behave autocatalyticaly,
leading to the extinction of the prey species and, consequently, the
predator population. Even before the information and quantitative
era, ancient day naturalists were able to recognize this puzzle.
Among them, Herodotus and Cicero perceived the persistence of prey
species in the face of adversity as a manifestation of divine power
and the creator's design (Cuddington, 2001). Herodotus' explanation
for the phenomenon was based on multiplication rates. In his words:
"... for timid animals which are a prey to others are all made to
produce young abundantly, so that the species may not be entirely
eaten up and lost; while savage and noxious creatures are made very
unfruitful." This explanation, clearly, can not stand on its own.
Even if Herodotus' statements about the reproduction habits of
certain species were true (he claims that the hare is able to
superfetete, i.e., to conceive while still pregnant; while the
lioness may use her womb only once, since the cub ruptures it with
his  claws during pregnancy), balancing two exponentially divergent
processes (multiplication and predation) is achievable only via
fine-tuning of the parameters. Moreover, that fine-tuning must be
robust against various kinds of noise. Indeed, as a rule, natural
prey-predator systems are always subject to noise: either some sort
of environmental stochasticity (e.g., years with more/less rain,
weather fluctuations, heterogenous habitat) or, even under fixed
conditions, some noise is introduced into the system due to
demographic stochasticity, related to the discrete nature of the
individual agents and the stochastic character of
birth-death-predation processes (Lande, 2003; Foley 1997).

In modern times, the course of  explanations has been changed,
becoming more deterministic and quantitative. The basic idea  relies
on the fact that when the prey population decays there is not enough
food for the predators and their population also diminishes, so both
populations oscillate around their mean values.  As pointed out by
Nicholson (1933), these oscillations are an intrinsic property of
interacting populations. If, for example, the density of a host is
above its steady value, it will be reduced by the parasite. However,
when the host reaches its steady density, the density of parasites
is above its steady value. "Consequently, there are more than
sufficient parasites to destroy the surplus hosts, so the host
density is still further reduced in the following generation ...
Clearly, then, the densities of the interacting animals should
oscillate around their steady value." This paradigm was formulated
mathematically,  using deterministic continuous-time partial
differential equations (PDE's), by Lotka and Volterra  (LV model)
(Lotka, 1920; Volterra 1931; Murray 1993). An analogous model with a
discrete time step was introduced for a parasitoid-host system by
Nicholson and Bailey (NB) (Nicholson and Bailey, 1935). Both models
[and their variants (May, 1978; Hassell and Varley, 1969; Crowley,
1981; Rosenzweig and MacArthur, 1963)] allow for a \emph{coexistence
fixed point}, i.e., a steady state that admits finite population of
the predator and the prey, and predict population oscillations
around this coexistence steady state. The prediction of population
oscillations is one of the major achievements of the deterministic
models. Its most well known example is century-long documentation of
oscillations in the Canadian lynx - Snowshoe Hare inhabitants of
Northern America (Elton, 1924; Stenseth et al., 1998).

However, beyond this success, Lotka,  Volterra and Nicholson
recognized that their scheme fails to solve the basic stability
problem, as it cannot account for  the sustainability of these
oscillations (Nicholson, 1933; Cuddington, 2001). If, under the
influence of demographic or environmental stochasticity, oscillation
amplitude increases, it should  eventually reach an extinction point
for  one of the species. But this is exactly the situation for
systems described by the Lotka-Volterra model: as oscillations of
any amplitude are allowed, any type of noise will eventually
increase their amplitude and the species will go extinct (see
section 2 below). For the Nicholson-Bailey map, the situation is
even worse: the coexistence fixed point is not only marginally
stable but unstable. Thus,  even without a continuous effect of
noise, any small perturbation should increase exponentially and
drive the system to extinction. Trying to maintain the coexistence
state in a NB host-parasite model is like trying to balance a cone
on its tip.

Facing the inability of the LV and NB  models to admit oscillation
sustainability,  one may feel a temptation to abandon them
altogether in favor of  another  formulation that supports an
attractive fixed point or (if oscillations  are required) a
limit-cycle/strange-attractor. While it is reasonable to assume that
some natural populations are described by such models, experimental
and field studies show that, quite generically, this instability is
a main characteristic of reality. In fact, for small systems and
well-mixed populations, experiments show oscillation growth and
extinction in a wide variety of interacting species (Gause , 1934;
Pimentel et al., 1963; Huffaker, 1958). Spatially-extended systems,
on the other hand, seem to support sustained oscillations, as
emphasized by field studies, experiments, (Kerr et al.,  2002; Kerr
et al., 2006; Luckinbill, 1974; Holyoak and Lawler, 1996) and
numerics (Wilson et al., 1993; Bettelheim et al., 2000; Washenberger
et al., 2006; Kerr et al., 2002; Kerr et al., 2006).

\begin{figure}
 \includegraphics[width=8cm]{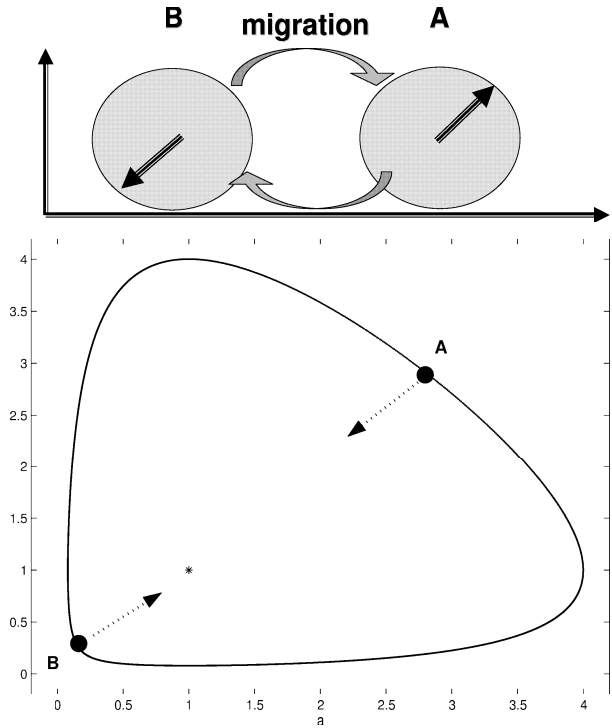}
 \caption{Population oscillation on two spatial patches coupled by
migration.
 If
 both patches desynchronize, one may find one of them (A) in the dense
population state
 and the other one (B)  in the dilute phase. Diffusion tends to decrease
population gradients, hence
 the whole system flows towards the coexistence fixed point, represented in
the lower panel by an asterisk.}
 \label{2patches}
\end{figure}

These  findings give an appeal to  Nicholson's (1933) old proposal
about migration induced stabilization, i.e., that
\emph{desynchronization} between weakly coupled spatial patches,
together with the effect of migration, stabilize the global
populations. To get the flavor of the mechanism, let us imagine a
metapopulation on two patches, where within each patch the
population oscillations are governed by, say, Lotka-Volterra
dynamics with  constant diffusion (i.e., constant per capita
migration rate) between these two patches. Clearly, as emphasized in
Figure \ref{2patches}, \emph{if} the oscillations on these two
patches desynchronized, e.g., if one of the patches is in a state of
large populations while the other is, at the same time, in a diluted
state, migration between patches pushes the whole system inward
toward the coexistence fixed point yielding sustained oscillations.
However, one should bear in mind that migration is a double edged
sword, as it tends to avoid population  gradients and leads to
synchronization. The acid test for Nicholson's proposal is thus as
follows: is the diffusion among patches weak enough to allow
noise-induced desynchronization, but at the same time strong enough
to stabilize desynchronized patches? If this is to be the case,  the
desynchronization-diffusion stabilization may work.

 Unfortunately, examination of this idea in many studies,
summarized in a recent review article (Briggs and Hoopes, 2004),
yields negative results. Generically, diffusion stabilizes the
homogenous manifold and different spatial patches get synchronized,
leading back to the well-mixed unstable dynamics (Crowley, 1981;
Allen, 1975; Reeve, 1990). At the end of the day, thus, we are back
where Herodotus left us more than 2500 years ago: we cannot explain
the sustainability of prey-predator systems, despite  their presence
since the beginning of life on earth. The understanding of
oscillation persistence is not only an intellectual challenge;
stabilization of such oscillations is considered to be a major
factor affecting species conservation and ecological balance (Earn
et al., 2000; Blasius et al., 1999).

The goal of this paper is to provide a \emph{generic} solution to
this puzzle, a solution that is independent of external assumptions,
such as space-time fluctuations or heterogenous migration patterns.
We will show that the basic ingredient that leads to
desynchronization is the dependence of oscillation frequency on
their amplitude (Amplitude Dependent Frequency, ADF), and will
survey the consequences and the implementation of this idea for
different systems and  types of noise. The basic quantitative and
qualitative tool presented here is a  simple toy model for coupled
oscillators. Apart from its usefulness in the analysis of the ADF
mechanism, this toy model works for all types of stabilizing
mechanisms presented so far, and serves as a prediction and
classification tool. As such, we suggest our coupled oscillators
system to be used as the standard -modeling technique for
metapopulation oscillations.

The present work is organized as follows: in the next section, the
stability problem will be introduced for a single patch system,
together with a short review of the experimental and numerical
results that support the instability for well-mixed populations. In
the following section, Nicholson's proposal - desynchronization
induced stability on spatial domains -  is endorsed, based again on
numerics and experimental results. The fourth section is the main
part of the paper, where a toy model is presented and analyzed and
the role of amplitude-dependent oscillations is clarified.  Next we
give a short critical review of previous solutions to the stability
problem, and show how to incorporate them all within the framework
of diffusively-coupled oscillators. We end up with some concluding
remarks, trying to sketch a classification scheme for observed
population dynamics in experimental and numerical situations.

\section{ Instability and extinction of well mixed populations}

We begin the demonstration of the instability problem  looking at
the  Lotka-Volterra predator-prey system, the paradigmatic model for
oscillations in population dynamics (Lotka, 1920;  Volterra, 1926;
  Murray, 1993). The model describes the temporal
evolution of two interacting populations: a prey ($b$) population
that grows with a constant birth rate $\sigma$ in the absence of a
predator (the energy resources consumed by the prey are assumed to
be inexhaustible), while the predator population ($a$) decays (with
death rate $\mu$) in the absence of a prey. Upon encounter, the
predator may consume the prey with  a certain probability. Following
a consumption event, the predator population grows and the prey
population decreases. For a well-mixed population, the corresponding
partial differential equations are:
\begin{eqnarray}\label{basic}
\frac{\partial a}{\partial t} &=& - \mu a + \lambda_1 a b \\
\nonumber \frac{\partial b}{\partial t} &=& \sigma b - \lambda_2 a b
\end{eqnarray}
where $\lambda_1$ and $\lambda_2$ are the relative increase
(decrease) of the predator (prey) populations due to the interaction
between species, correspondingly.

\begin{figure}
 \includegraphics[width=8cm]{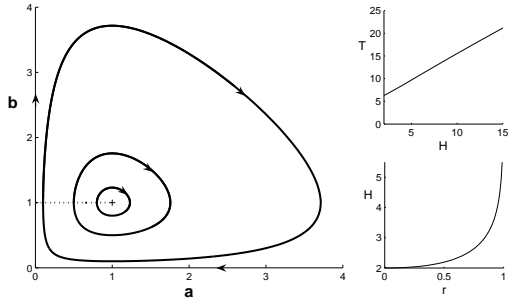}
 \caption{The Lotka-Volterra phase space (left panel) admits a
 marginally stable fixed point surrounded by close trajectories (three of
these
 are
 plotted). Each trajectory corresponds to a single $H$ defined in Eq.
 (\ref{H}), where $H$ increases monotonically  along the (dashed) line
 connecting the center with the $a=0$ wall, as shown in the lower
 right panel. In the upper right  panel, the period of a cycle $T$ is
 plotted against $H$, and is shown to increase almost linearly from
 its initial value $T = 2 \pi / \sqrt{\mu \sigma}$ close to the
 center.}
 \label{phasespace}
\end{figure}

The system admits two unstable fixed points: an absorbing state
$a=b=0$ and the state $a=0, \ \ b = \infty$. There is one
coexistence, marginally stable fixed point at $\bar{a}  =
\sigma/\lambda_2, \ \ \bar{b} = \mu/\lambda_1$. Local stability
analysis yields the eigenvalues $\pm
 i \sqrt{\mu \sigma}$ for the stability matrix. Moreover, even
 beyond the linear regime there is neither convergence nor
 repulsion. Using logarithmic variables $z = ln(a), \  q = ln(b)$
 eqs. (\ref{basic}) take the canonical form $\dot{z} = \partial
 H/\partial q, \ \ \dot{q} = -\partial
 H/\partial z$, where the conserved quantity $H$ (in the $ab$
 representation) is:
 \begin{equation}\label{H}
 H = \lambda_1 b + \lambda_2 a - \mu \ ln(a) - \sigma \  ln(b).
 \end{equation}
The phase space is thus  segregated into a collection of nested
one-dimensional trajectories, where each one is characterized by a
different value of  $H$, as illustrated in Figure \ref{phasespace}.
Given a line connecting the fixed point to one of the "walls" (e.g.,
the dashed line in the phase space portrait, Figure
\ref{phasespace}), $H$ is a monotonic function on that line, taking
its minimum $H_{min}$ at the marginally stable fixed point (center)
and diverges on the wall. It turns out that all the important
features related to the instability and the synchronization depend
only on the topology of the phase space, and the actual values of
the growth, death and predation parameters are not important. Thus,
without loss of generality, we employ hereon (unless otherwise
stated)  the symmetric parameters $\mu = \sigma = \lambda_1 =
\lambda_2 =1$.  The corresponding phase space, along with the
dependence  of $H$ on the distance from the center and a plot of the
oscillation period vs. $H$, are represented in Figure
\ref{phasespace}.

\begin{figure}
 \includegraphics[width=7cm]{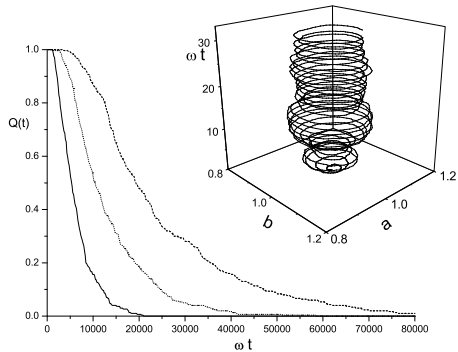}
 \caption{The survival probability $Q(t)$ is plotted versus time for
a single patch noisy LV
 system. Eqs. (\ref{basic}) (with the symmetric parameters) were integrated
numerically (Euler integration
 with time step $0.001$), where the initial conditions are at the fixed
point $a=b=1$. At each time step, a small random number
 $\eta(t) \Delta t$ was added to each population density, where $\eta(t)
\in [-\Delta,\Delta]$. A typical phase space trajectory, for
 $\Delta = 0.5$, is shown in the inset. The system "dies" when the
trajectory hits the walls $a=0$ or $b=0$. Using 300 different noise
histories, the
 survival probability is shown here for $\Delta = 0.5$ (full line), $\Delta
= 0.3$ (dotted line) and $\Delta = 0.25$ (dashed line).
 Clearly, the survival probability decays exponentially at long
 times, $Q(t) \sim exp(-t/\tau)$, as expected for a random walk
 with absorbing boundary conditions, $1/ \tau$ scales with
 $\Delta^2$.}
 \label{fig2}
\end{figure}

Given the integrability of that system, the effect of noise is quite
trivial: if $a$ and $b$ randomly fluctuate in time, the system
wanders between trajectories, thus performing some sort of random
walk in  $H$  with "repelling boundary conditions" at $H_{min}$ and
"absorbing boundary conditions" on the walls (as negative densities
are meaningless, the "death" of the system is declared when the
trajectory hits the zero population state for one of the species,
i.e., when at least one of the species becomes extinct).  Thus,
under the influence of noise, thus, the problem is reduced to that
of a random walk with an absorbing trap [a first passage problem
(Redner, 2001)]. While the first passage problem is characterized by
power law decay, here the topology of the orbits forces the
trajectory back to the vicinity of the absorbing walls, and hence
the decay is exponential. An example of a phase space trajectory for
a single patch  noisy LV system is shown in the inset of  figure
\ref{fig2}, and the average growth of the oscillation amplitude is
clearly seen.  The survival probability $Q(t)$ (the probability that
a trajectory does not hit the absorbing walls within  time $t$) is
shown for different noise amplitudes in figure \ref{fig2}.

At this point it is instructive to clarify the types of noise to be
used throughout this paper. The first and the most simple is the
additive noise, i.e., random  addition or subtraction of population
density at certain time steps. In the real world, this corresponds
to migration of small parts of the population in and out of the
considered system. As the implementation of an additive noise in
Langevin type equations is relatively simple, we use it as the
"noise of choice" for the present work.

Indeed, there are two common sources for time-dependent randomness
in ecological systems. One source  is the environmental
stochasticity, i.e., fluctuations in the environmental conditions
that may change the reaction and interaction  parameters. The growth
rate of the prey, for example, should be taken not as a constant
$\sigma$ but as a fluctuating quantity around some average, $
\langle \sigma \rangle + \delta \sigma (t) $, where $\langle  \cdots
\rangle$ stands for average over time. Another source of noise is
the \emph{intrinsic} demographic stochasticity. While Eqs.
(\ref{basic}), for example, describe  continuous populations sizes
$a(t)$ and $b(t)$ that may take any value, the  actual number of
individuals in a community should be an integer (Lande et al.,
2003). Indeed, Partial differential equations like (\ref{basic})
only describe the dynamic of the average population and are derived
from the exact master equation as a "mean field" approximation,
neglecting higher correlations and approximating, e.g., $\langle
a(t) b(t) \rangle \sim \langle a(t) \rangle \langle b(t) \rangle$.
The actual dynamics of the system, described by the master equation,
is not deterministic but stochastic, and this implies that noise of
order $\sqrt{N}$ is applied to any population of $N$ individuals.
While the relative amplitude of that noise decays like $1/\sqrt{N}$
for  large populations, it is still of importance if there is no
attractive manifold for the deterministic dynamics (Kessler and
Shnerb 2006), such as in the LV or NB models. The effect of
demographic stochasticity on a single path LV system is very similar
to that of additive noise, as indicated by the simulation results
obtained using Gillespie's event driven algorithm (Gillespie, 1977).

In the current work, additive noise is used in order to model
multiplicative noise such as  demographic stochasticity. This
procedure may be justified using a self-consistency argument:
 we want to present a mechanism that stabilizes the
population oscillations, such that the number of individuals in,
say, the predator population, does not deviate strongly from its
average value. If this is the case, the ${\cal O} (\sqrt{N})$ noise
 amplitude does not change so much along the orbit and the system "feels"
constant (additive) noise. The noise amplitude of environmental
stochasticity is proportional to the population size [the term
$\delta \sigma (t)$ presented above multiplied by the prey
population in (\ref{basic})], thus it is also approximated by an
additive noise with  a constant amplitude if the oscillations are
not too large.

\begin{figure}
 \includegraphics[width=7cm]{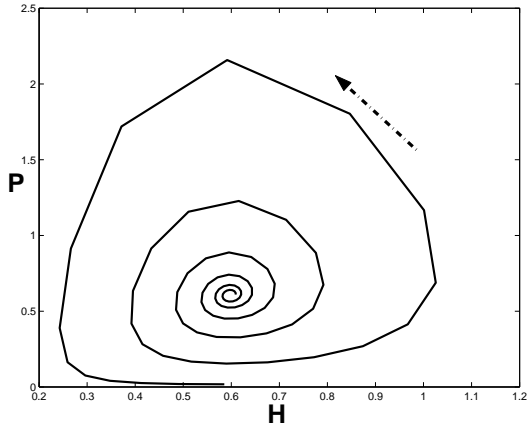}
 \caption{Nicholson-Bailey dynamics. The map (\ref{ni}) is iterated 90 times
where the initial deviation from the fixed point is 0.001.
 The parameters are $\sigma = 1.2$, $\lambda = 0.3$, $c  = 6.1$.   }
 \label{NBsingle}
\end{figure}

While the LV system is marginally stable and is driven to extinction
by the noise,  the situation for the host-parasitoid model of
Nicholson and Bailey is even worse.  In their model, the host
population $H$ and the parasite population $P$ satisfy the map
(Nicholson and Bailey, 1935):
\begin{eqnarray} \label{ni}
H_{t+1} = \sigma H_t e^{-\lambda P_t}, \nonumber \\
P_{t+1} = c H_t (1-e^{-\lambda P_t})
\end{eqnarray}
where $\sigma >1$ is the growth factor of the host in the absence of
a parasite.  The system admits a  coexistence fixed point at:
\begin{eqnarray}
H = \frac{ln \  \sigma}{\lambda}, \nonumber \\
P =  \frac{\sigma ln  \ \sigma}{c \lambda (\sigma-1)}.
\end{eqnarray}
and small deviations from that fixed point grow exponentially with
time, yielding a prompt extinction of one of the interacting species
(see Figure \ref{NBsingle}). Here the effect of noise is negligible:
any infinitesimal deviation from the fixed point renders the system
extinction-prone.

\section{Spatial structure and sustained oscillations}

As concluded in the last section, a well-mixed population that obeys
Lotka-Volterra or Nicholson-Bailey dynamics is extinction prone:
oscillation amplitude grows in time until one of the species goes
extinct. The only difference between the two models is the way this
extinction is approached: while an NB system is unstable and
oscillations grow in time exponentially even without noise, the LV
model admits neutral oscillations and  approaches extinction only
due to the effect of stochasticity.

One may thus suggest that something is wrong in the modeling of
interspecific interaction. While LV and NB are clearly the simplest
descriptions of the corresponding systems, maybe in nature some more
complicated processes take place, and more accurate mathematical
models should use equations that support an \emph{attractive
manifold}, i.e., a phase space region that "attracts" the
trajectories. In particular, one may write a model that supports a
stable fixed point,  limit cycle (Kolmogorov, 1936; May, 1972) or
strange attractor. In the first case,  after a short period of
oscillations the system settles into a stable state of constant
populations; in the second, it will converge to fixed amplitude
oscillations; and in the third scenario (strange attractor), it will
wander chaotically between a set of points confined to some phase
space region.  Any of these solutions is robust against small noise,
as the dynamic is dissipative, small fluctuations flow back into the
attractive manifold.

However, a series of experimental results shows that, indeed, for
small sample sizes,  prey-predator or host-parasitoid systems
actually flow to extinction. From  the seminal work of  Gause on the
Didinium - Paramecium system (Gause, 1934), through  Huffaker's
orange experiment (six spotted mite - Typhlodromus ) (Huffaker,
1958), and Pimintel's wasp-fly setup (Pimentel et al., 1963); all of
these experiments demonstrated that population dynamics result in
extinction of one of the species.

Of particular interest is a new series of experiments dealing with
the cyclic, rock-paper-scissors dynamic (Kerr et al., 2002; Kirkup
and Riley, 2004; Kerr et al., 2006) . In the first set of
experiments, colicin-producing strains (C) of E. Coli were
interacting with colicin-sensitive  (S) and colicin-resistent
strains (R); while C kills S, S outcompetes  R and R wins against C.
It turns out that for well-mixed systems, both \emph{in vitro} (Kerr
et al., 2002) and \emph{in vivo} (Kirkup and Riley, 2004),
fluctuations drive two of the three species to extinction via the
same mechanism of "random walk" among naturally stable orbits
(Reichenbach et al., 2006).  In the second experiment (Kerr  et.
al., 2006) E. Coli and its parasitoid (phage) were subjected to
controlled migration. Again, in the "well-mixed" regime, when all
the biological materials were mixed each 12 hours, extinction took
place after a very short time.

On the other hand, the spatial structure and, in particular, spatial
migration have been shown to be of crucial importance in the
sustainability of interacting populations. The famous field studies
about natural prey-predator systems [like the Canadian Lynx -
Snowshoe Hare (Elton, 1924; Stenseth et. al., 1998) data from Hudson
Bay company, the Moose-Wolf coexistence in Isle Royal (Wilmers, et
al., 2006), and the biological control of the Prickly pear cactus by
the moth cactoblastis cactorum in eastern Australia (Freeman, 1992)]
indicate that, in spatially-extended systems, the population
oscillations are either negligible or finite,  ensuring persistence.
Luckinbill's (Luckinbill, 1974) experiment, using again the Didinium
- Paramecium system, shows that the system's lifetime (time until
extinction of one of the species) grows almost exponentially with
its linear size. Similar observations were made by Holyoak and
Lawler  (1996), and, as already mentioned, by the series of
experiments dealing with rock-paper-scissors dynamics (Kerr et. al.,
2002; 2006) , where slow mixing yields finite fluctuations and
promotes survival. There are also many numerical experiments on
stochastic models (like the individual-based models, subject to
demographic stochasticity) that show stable oscillations for large
samples(Wilson, et al., 1993; Bettelheim et al., 2000; Washenberger
et al.,  2006 ;Kerr et. al., 2002; 2006).

All in all, much evidence supports the idea that interacting species
dynamics are inherently unstable on a single patch, and that the
stability observed in nature is related to the fact that natural
systems are spatially-extended. This idea may be implemented in an
extinction-recolonization scenario (Taylor, 1990), but in general
this mechanism works for any system that supports desynchronization.
As pointed out by Nicholson (1933), \emph{if} the system
desynchronizes, the migration between patches will induce an inward
flow toward the coexistence fixed point and stabilize the amplitude
of oscillations. The main puzzle, thus, is reduced to the
identification of the conditions under which desynchronization takes
place. It turns out, however, that generically the diffusion tends
to lock the system to its synchronized state, stabilizing the
homogenous manifold such that the whole system acts like a single
patch, rendering unbounded oscillations (Briggs and Hoopes , 2004).

The condition for  desynchronization in diffusively coupled patches
have  been examined in many studies and the main results, summarized
in a recent review article (Briggs and Hoopes , 2004), are as
follows:

\begin{itemize}

\item For any network of $N$ patches, if the migration between
patches is symmetric or almost symmetric (i.e., the diffusion of the
prey and the predator are, more or less, the same), there is no
diffusion-induced instability, and the homogenous manifold is stable
(Crowley, 1981; Allen, 1975; Reeve, 1990). Thus, the effect of
migration alone does not cure the instability problem.

\item The system may become desynchronized when in the presence
of spatial heterogeneity, e.g.,
 where the reaction parameters vary on different spatial
patches (Murdoch and Oaten, 1975; Murdoch et al., 1992; Hassell and
May, 1988). In that case, the intrinsic dynamic at any localized
patch takes place on different time scales for the same
concentrations, so diffusion fails to synchronize different patches.
This mechanism may be generalized to include not only "quenched"
heterogeneity but also environmental stochasticity, i.e.,where the
reaction parameters are subject to spatio-temporal fluctuations
(Crowley, 1981 ; Reeve, 1990; Taylor, 1998) .

\item Spatial heterogeneity may be introduced into the model in a
more sophisticated way: one may define a homogenous dynamics in all
patches with an underlying nonlinear process that supports
spontaneous symmetry breaking and pattern-formation. These patterns,
in turn, will control the reaction parameters to yield
desynchronization.

\item  Diffusion-induced instability may occur if the migration rate
of the predator is much smaller than that of the prey, particulary
if the prey migration rate is zero [Jansen, 1995; Abta et al. (2006
b)].

\end{itemize}

While spatial heterogeneity, environmental stochasticity, and
different migration rates for different species are indeed
characteristics of many natural systems, these mechanisms are not
generic and it seems that they cannot explain the experimental
results that show persistence in "clean"  extended systems, like
(Luckinbill, 1974) and (Kerr et al.; 2002; 2006). Accordingly, it
seems  that the combined effect of noise and diffusion is a
necessary precondition for population stabilization. However, up
until now, the qualitative nature of the "missing" underlying
mechanism has remained obscure, and no theoretical framework that
allows for quantitative prediction has been presented.

This  theoretical gap has been addressed recently (Abta et al., 2006
a) using a toy model for coupled oscillators presented in the next
section. The main new ingredient emphasized by the proposed model is
the \emph{dependence of frequency on the oscillation amplitude},
reflected by the gradient of the angular velocity along the radius
$\omega'(r)$. The instability induces desynchronization iff the
small, noise-induced differences between patches are amplified  by
the frequency gradient such that the "desynchronization parameter"
$\langle \phi^2 \rangle$ acquires a finite value, leading to
"restoring force" toward the origin of the homogenous manifold.

Moreover, it turns out that our toy model may also reproduce the
other mechanisms suggested in the past, and yield simple analytical
results together with qualitative and quantitative predictions. In
the fifth section, these generalizations are presented. Thus, our
toy model allows for a comprehensive typification of different
mechanisms for sustained oscillation on diffusively coupled patches,
and yields simple criteria for classification of experimental
results and identification of the stabilizing mechanism.

\section{Two patch system}

To consider the effect of spatial structure, let us begin with the
simplest case, i.e., two patch system connected diffusively as in
Figure \ref{2patches}. One should bear in mind, though, that any
finite system that admits an absorbing state  and is subject to
additive noise must eventually reach extinction. Under the influence
of noise, any stable state becomes only metastable, as some rare
configuration of the noise will push one of the species to
extinction. The appearance of sustained oscillations, i.e., of an
attractive manifold,  manifests itself in finite system only in its
lifetime  before it hits the absorbing state. This lifetime, $\tau$,
is known in the literature as  the "intrinsic mean time to
extinction" (Grimm and Wissel, 2004), and if a metastable state
appears,  it increases exponentially with the "strength" of
attraction (the Lyapunov exponent) of that manifold. If the
stability of  a system increases with the number of patches, one may
expect the lifetime of the system to diverge with the number of
patches, yielding real stability for an infinite system. The best
example of this phenomenon - a metastable state in a small system
becoming stable as its size grows - is the logistic growth with
demographic stochasticity, where the transition from extinction to
proliferation belongs to the directed percolation equivalence class
(Grassberger, 1982;  Janssen, 1981). A similar transition, belonging
to the same universality class, was observed recently for coupled
limit cycles with demographic stochasticity (Mobilia, et al,  2006).
Thus we are not trying to show that the two patch system is really
stable; exponential growth in the lifetime of the system is a clear
indication of the appearance of an attractive manifold that yields
real stability in the infinite size limit.

The simplest example is the LV system on two patches:
\begin{eqnarray}\label{two}
\frac{\partial a_1}{\partial t} &=& - \mu a_1 + \lambda_1  a_1 b_1 +
D_a (a_2-a_1) \nonumber \\   \frac{\partial a_2}{\partial t} &=& -
\mu a_2 + \lambda_1   a_2 b_2 + D_a (a_1-a_2)\\  \nonumber
\frac{\partial b_1}{\partial t} &=& \sigma b_1 - \lambda_2 a_1  b_1
+ D_b (b_2 - b_1) \\ \nonumber \frac{\partial b_2}{\partial t} &=&
\sigma b_2 - \lambda_2  a_2 b_2 + D_b (b_1 - b_2).
\end{eqnarray}

\begin{figure}
 \includegraphics[width=9cm]{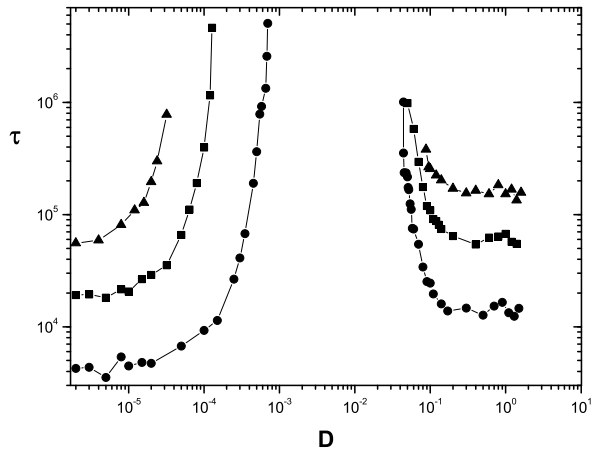}
 \caption{The typical persistence time as a function of the diffusion
rate for different levels of noise.
 The values of $\tau$ were gathered from survival probability plots (like
those in Figure  \ref{fig2}) and are displayed here
 for the two-patch system. One sees that the value of $\tau$ grows very
rapidly (even faster than exponentially)
 with the migration rate for small diffusion values, and decays with $D$
for large diffusivities. Data is shown for different
 noise intensities $\Delta = 0.3$
  (triangles), $0.5$ (squares) and $1.0$ (circles).}
  \label{fig3}
\end{figure}

The time evolution of that system, with additive noise, equal
diffusivities, $D_a=D_b \equiv D$, and symmetric reaction rates  is
obtained through Euler integration. In the limit $D=0$ the patches
are unconnected; thus, starting from the homogenous fixed point $a_1
= a_2 = b_1 =b_2 =1$, the single patch situation  still holds and
the system hits the absorbing walls after a characteristic,
noise-dependent, time. In the opposite limit, $D=\infty$, the system
sticks to the invariant manifold and acts like a single patch (with
modified noise and interaction parameters), again performing  a
random walk in the invariant manifold. However, between these two
extremes, there is a region where the combined effect of diffusion
and noise stabilizes a finite region within the invariant manifold.
For any noisy two-patch system $Q(t)$, histograms (like those shown
for a single patch in Figure \ref{fig2}) were used to extract the
typical decay time ["intrinsic mean time to extinction" (Grimm and
Wissel, 2004)]  $\tau(D, \Delta)$ by fitting its tail to exponential
decay $exp(- t/\tau)$. In Figure \ref{fig3}, $\tau(D, \Delta)$ is
plotted against $D$ for different noise amplitudes, and is shown to
increase (faster than exponentially) with $D$ ($1/D$) as $D$
approaches zero (infinity). Evidently,  for intermediate
diffusivities, an attractive manifold appears in phase space, with a
Lyapunov exponent that grows (faster than linearly) with the
diffusion constant.

\subsection{The Coupled oscillator model and  LV system}

\begin{figure}
 \includegraphics [width=8cm] {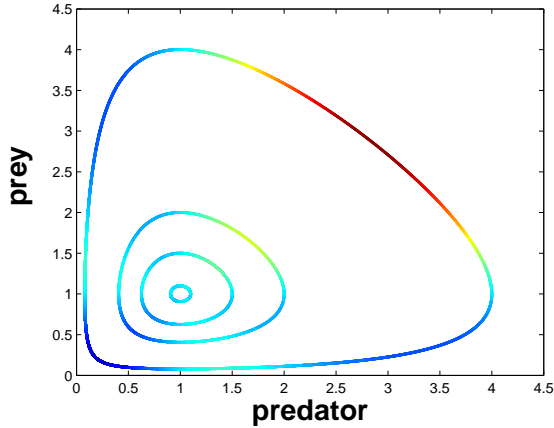}\\
 \caption{The angular velocity along some orbits of the  Lotka-Volterra
 dynamic. Fast regions are marked in red, slow regions in blue.
  Clearly, the dynamic is slowest when the populations of both species are
diluted, and fastest along the
  dense region in the upper-right "shoulder." Note that the velocity
gradient along an orbit increases with $H$. }
 \label{ang}
\end{figure}

Although both the Lotka-Volterra  and  Nicholson-Bailey  models are
oversimplified with respect to the complexity of interspecies
interactions in the natural world, they are still too complicated to
allow for an understanding of the  stabilization mechanism when
migration and noise interfere. The main obstacle (as will become
clear later) is that the radial velocity along a trajectory depends
not only on the amplitude of oscillations  (related to the conserved
quantity $H$), but also on the location along a trajectory, i.e., on
the azimuthal angle between the point along the trajectory and the
fixed point $\theta$ (see Figure \ref{ang}).  In order to clarify
the origin of the stable cycles, let us introduce a toy model that
imitates the main features of the real systems. Although that model
does not allow for an absorbing state, it captures the basic
mechanism for stabilization of spatially-extended systems in the
presence of noise.

The toy model deals with the phase space behavior of
diffusively-coupled oscillators, where the angular frequency depends
on the radius of oscillations. With additive noise, the Langevin
equations take the form,
\begin{eqnarray}\label{oscillators}
\frac{\partial x_1}{\partial t} &=& \omega(x_1,y_1) y_1 + D_1
(x_2-x_1) + \eta_1(t) \nonumber
\\ \nonumber \frac{\partial x_2}{\partial t} &=& \omega(x_2,y_2) y_2+
D_1 (x_1-x_2)+ \eta_2(t)  \\ \frac{\partial y_1}{\partial t} &=&
-\omega(x_1,y_1) x_1 + D_2 (y_2 - y_1)+ \eta_3(t) \\ \nonumber
\nonumber \frac{\partial y_2}{\partial t} &=& -\omega(x_2,y_2) x_2 +
D_2 (y_1 - y_2) + \eta_4(t),
\end{eqnarray}
where all the $\eta$-s are taken from the same distribution.  If the
angular frequency is location independent, $\omega(x,y) = \omega_0$,
the problem is reduced to coupled \emph{harmonic} oscillators, a
diagonalizable linear problem that admits two purely imaginary
eigenvalues in the invariant, homogenous manifold. With noise, the
random walk on that manifold is independent of the motion in the
fast manifold and the radius of oscillation diverges with the square
root of time.

 Now let us define the oscillation radius for each
patch, $r_i = \sqrt{x_i^2 + y_i^2}$ for $i=1,2$, and assume that the
angular frequency depends only on that radius and is
$\theta$-independent [$\theta_i \equiv arctg(y_i/x_i)$]. With that,
the total phase $\Phi = \theta_1 + \theta_2$ decouples and the
3-dimensional phase space motion is dictated by the equations (we
take $D_1 = D_2 = D$ and define $\phi = \theta_1 - \theta_2$, $R
\equiv r_1+r_2$, $r \equiv r_1 -r_2$):
\begin{eqnarray}\label{3d}
 \dot{R} &=& -2D sin^2(\phi/2)R + \tilde{\eta}_R  \\
 \dot{r} &=& -2Dcos^2(\phi/2)r + \tilde{\eta}_r  \\
 \dot{\phi} = -2&D&
 \left( \frac{R^2+r^2}{R^2-r^2} \right) sin \phi +
\omega(r_2) - \omega(r_1) +  \left( \frac{\tilde{\eta}_{1}}{r_1} -
\frac{\tilde{\eta}_{2}}{r_2} \right). \label{7}
\end{eqnarray}
As before, all the $\tilde{\eta}$-s are taken from the same
distribution. Note that $\phi$ represents the phase
desynchronization between patches while $r$ is the amplitude
desynchronization.

\begin{figure}
 \includegraphics[width=8cm]{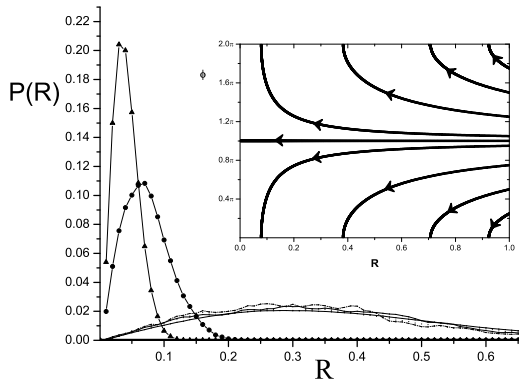}
 \caption{Histograms showing the probability to be at  distance $R$ from
the origin as a function of $R$, for
 two coupled noisy oscillators, where $\omega = 1+ \alpha r$  with $D=0.01$
with various values of noise strength $\Delta$ and
 angular velocity gradient $\alpha$. As expected, the phase space
confinement is proportional to $\alpha$, from $\alpha = 1$
(triangles)
 to $\alpha = 0.5$ (circles) to $\alpha = 0.1$ (solid line), all for the
same level of noise $\Delta = 0.1$. On the other hand,
 as predicted by the linear analysis close
 to the invariant manifold, the confinement is noise-independent, and the
three solid lines corresponding to
 different levels of noise ($\Delta = 0.1,0.5,1$) with the same $\alpha=0.1$
almost coincide. The inset shows the flow lines
 on the $r=0$ plane. The invariant manifold $\phi =0$ is stable, but there
is a "maximally desynchronized" unstable orbit
 converging to the center at $\phi = \pi$. If the expectation value of
$\phi^2$ deviates from zero, there is an effective restoring force
 toward the center, and the noise-induced random walk on the $\phi =0$
manifold is bounded. }
 \label{fig4}
\end{figure}

Eqs. (\ref{3d}) clarify the role of  phase desynchronization as the
stabilizing mechanism. The dynamics in  the homogenous ($R$)
manifold look very much like that of an overdamped harmonic
oscillator  in noisy environment,
\begin{equation}
\dot{z} = -kz+ \eta(t),
\end{equation}
a well-known system [Ornstein-Uhlenbeck process, (Gardiner, 2004)]
that admits the steady state Boltzman distribution $P(z) \sim exp(-k
z^2/\Delta^2)$. However, if the phases of these two patches
synchronize and the expectation value of $\phi^2$ vanishes, so does
the "spring constant" of the oscillator. Without that restoring
force, the motion on the $R$ manifold is a simple random walk, so
the oscillation amplitude grows indefinitely. Phase ($\phi$)
desynchronization, thus,  is the crucial condition for
stabilization. This feature is stressed in the inset of Figure
\ref{fig4}, where the flow lines of the deterministic dynamics in
the $R-\phi$ plane are sketched: the line $\phi = 0$ is marginally
stable, but any deviation leads to inward flow.

Close to the invariant manifold, when $\phi$ and $r/R$ are much
smaller than one, the amplitude desynchronization $r$ is solvable.
Neglecting corrections of order $\phi^2$,
\begin{equation}
\dot{r} = -2Dr+\eta, \end{equation} which is again an equation for
an overdamped harmonic oscillator, so
\begin{equation}
\label{cc} P(r) \sim exp(-Dr^2/\Delta^2) \end{equation}
 and the
$r^2$ typical fluctuation around zero is of order $\Delta^2/D$.
However, the amplitude desynchronization factor $r$ does not appear
in (\ref{3d}), and the stabilization is determined only by the
phase. Accordingly, the system  supports an attractive manifold iff
the amplitude desynchronization yields phase desynchronization.

Another piece of information may be gathered from the harmonic
limit, $\omega(r)  = \omega_0$. Here there should be no phase
synchronization, as we already diagonalized the linear equation and
find no restoring term in the homogenous plane. Looking at Eq.
(\ref{7}) with $\omega(r_1) - \omega(r_2) = 0 $ one concludes, thus,
that the rightmost (noise) term in (\ref{7}) is irrelevant. The
dependence of the frequency on the  amplitude (i.e., the dependence
of $\omega$ on $r$) should be the factor that allows the translation
of amplitude desynchronization into a phase desynchronization.
Intuitively, when two patches with different oscillation amplitudes
move with different angular velocities, this immediately yields
phase differences.

\begin{figure}
 \includegraphics[width=8cm]{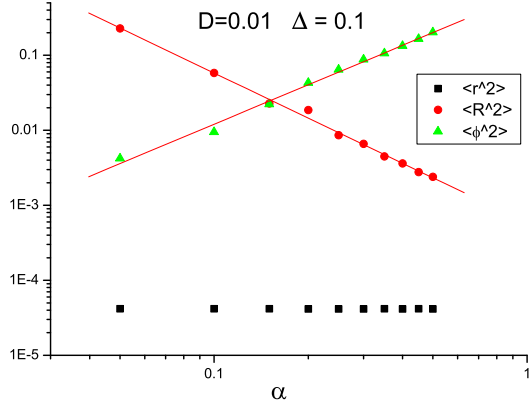}
 \caption{ $\langle \phi^2 \rangle$ (triangles) $\langle r^2 \rangle$
(squares) and $\langle R^2 \rangle$ (circles)
 as a function of $\alpha$, for two coupled oscillators where $\omega = 1+
\alpha r$, $D = 0.01$
  and $\Delta = 0.1$. Clearly the value of  $\langle r^2 \rangle$
   is almost independent of the frequency gradient $\alpha$. $\langle R^2
\rangle$ is fitted to $\alpha^{-1.99}$, in agreement with Eq.  (
   \ref{pR}). $\langle \phi^2 \rangle$ scales like $\alpha^{1.74}$,
   close to the prediction of (\ref{php}).
   }
 \label{omp}
\end{figure}

\begin{figure}
 \includegraphics[width=8cm]{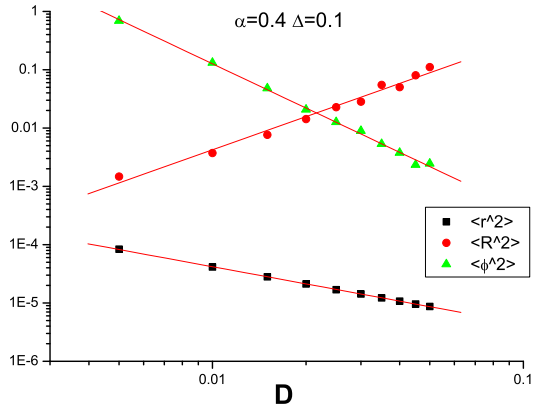}
 \caption{ $\langle \phi^2 \rangle$ (triangles) $\langle r^2 \rangle$
(squares) and $\langle R^2 \rangle$ (circles)
 as a function of the migration rate $D$, for two coupled oscillators where
$\omega = 1+ \alpha r$, $\alpha = 0.4$
  and $\Delta = 0.1$. As predicted,   $\langle r^2 \rangle$ is inversely
proportional to $D$ (best fit to $D^{-0.98}$). $\langle R^2 \rangle$
is fitted to $D^{-1.88}$,
  [predicted exponent is (-2)] and  $\langle \phi^2 \rangle$ scales like
$D^{-2.52}$ (predicted value
  -3).
   }
 \label{dp}
\end{figure}

\begin{figure}
 \includegraphics[width=8cm]{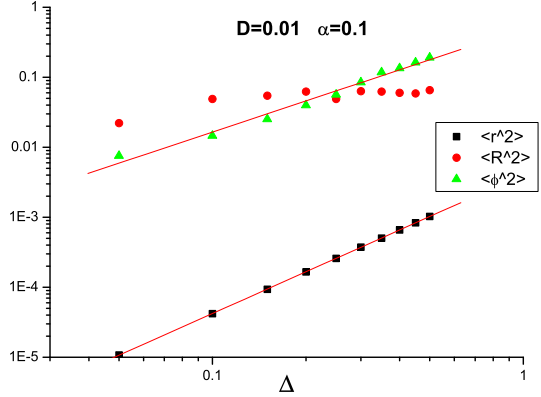}
 \caption{ $\langle \phi^2 \rangle$ (triangles) $\langle r^2 \rangle$
(squares) and $\langle R^2 \rangle$ (circles)
 as a function of the noise $Delta$, for two coupled oscillators where
$\omega = 1+ \alpha r$, $\alpha = 0.1$
  and $D = 0.01$. As predicted,   $\langle r^2 \rangle$ is proportional to
$\Delta^2$ (best fit to $\Delta^{-1.98}$). $\langle \phi^2 \rangle$
is fitted to
  $\Delta^{1.47}$,
  (predicted exponent is 2) and  $\langle R^2 \rangle$ clearly
  saturates at large noise levels.
   }
 \label{delp}
\end{figure}

\begin{figure}
 \includegraphics[width=9cm]{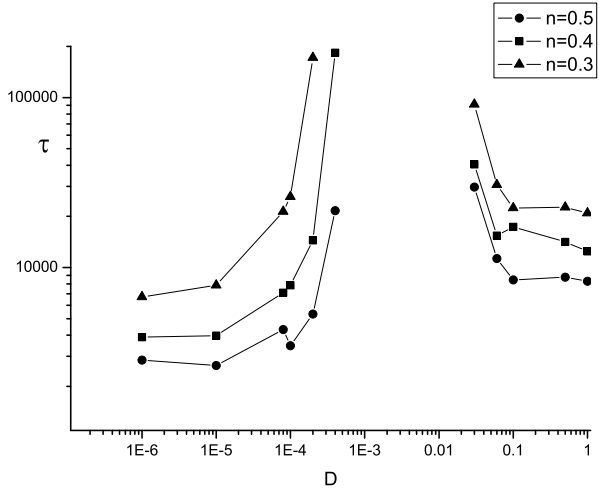}
 \caption{The typical persistence time as a function of the diffusion
rate for different levels of noise for the coupled oscillator model.
The "extinction" was declared when the amplitude of oscillations for
one of the patches exceeded unity.
 The values of $\tau$ were gathered from survival probability plots (like
those in Figure
  \ref{fig2}). One clearly observes the
  similarity with Figure \ref{fig3}.  Data is shown for different
 noise intensities $\Delta = 0.3$
  (triangles), $0.4$ (squares) and $0.5$ (circles). Note that
  while the expectation value of $R^2$ is noise independent, the
  noise still controls the probability to rare fluctuations that
  leads to extinction.}
  \label{avishag}
\end{figure}

\begin{figure}
 \includegraphics[width=8cm]{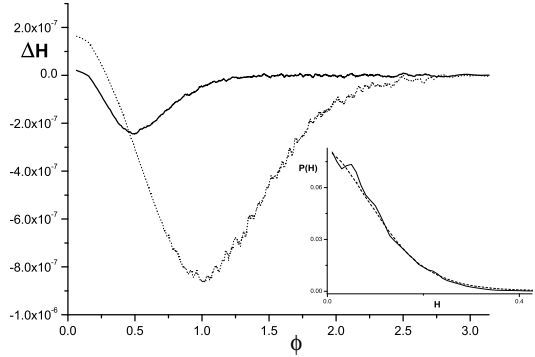}
  \caption{The average $\Delta H$ at an elementary time step (0.001 of a
unit time) as a function of the angle $\phi$ between the
  patches. While a simple phase space random
 walk yields, on average, positive $\Delta H$, this property is shown here
to hold only for small $\phi$. At larger
 angles,
 the diffusion between patches forces the system toward the center and the
average $\Delta H$ becomes negative.
 Results are shown for $\Delta = 0.1$ (full line)
 and $\Delta = 1$ (dashed line). The inset shows the probability
distribution function for $H$ at these two noise levels.}
 \label{fig5}
\end{figure}

 Without the explicit noise term, Eq. (\ref{7}) may be written as,
\begin{equation}
\dot{\phi} = -2D
 \left( \frac{R^2+r^2}{R^2-r^2} \right) sin \phi - r \frac{\partial
 \omega}{\partial r} \approx -2D \phi -r \frac{\partial
 \omega}{\partial r},
\end{equation}
where the approximation is valid close  to the invariant manifold.
Again we face an overdamped harmonic oscillator, where now the
source of noise is the $r$ fluctuations (obeying  the Boltzman
statistics).  With that,
\begin{equation} \label{php}
\langle \phi^2 \rangle \sim  \frac{(\omega^{'}(r))^2 \Delta^2} {
D^3},
\end{equation}
 may be plugged  into (\ref{3d}) to yield:
\begin{equation}
 \dot{R} = -2D sin^2 \left(\frac{ \phi}{2} \right) R + \tilde{\eta}_R
\approx \frac{ -D \langle \phi^2 \rangle R} {2}  + \tilde{\eta}_R.
 \end{equation}
Now the "restoring force" along the R coordinate (the invariant
manifold) is finite, and,
\begin{equation} \label{pR} \langle R^2 \rangle \sim  \frac{D^2}
{[\omega^{'}(r)]^2}.
\end{equation}
 This radius of stable oscillations diverges as
$D \to \infty$, as expected. The small $D$ instability (decoupled
patches) manifest itself in the divergence of $r^2$ as $D \to 0$.
Surprisingly, since both the restoring force and the noise in the
invariant manifold are proportional to $\Delta^2$, the expected $R$
distribution has to be \emph{noise-independent} at that limit, as
demonstrated in Figures \ref{fig4} and \ref{fig5}. A more careful
examination of the predictions about the expectation value of the
three parameters (Eqs. \ref{pR}, \ref{php} and the Boltzman
distribution for amplitude desynchronization) is presented in Figs.
\ref{dp}, \ref{omp}, and \ref{delp}. The small deviations from the
predicted exponents are related to the failure of the approximations
used far from the invariant manifold. Decreasing the noise level,
these exponents approach their predicted values. Figure
\ref{avishag} shows the $D$ dependence of the lifetime for the
coupled oscillator model; its agreement with figure \ref{fig3} is
evident.

Another direct manifestation of the coupled oscillator predictions
in the LV system is presented in Figure \ref{fig5}. Again,  for
 "intermediate"  migration (e.g., $D=0.01$), the average
distance from the origin saturates, while the chance to find the
system at large $H$ becomes exponentially small, as illustrated in
Figure \ref{fig5}. In agreement with the results of the  toy model,
the flow toward the center is correlated with the phase
desynchronization, leading to stabilization of oscillations at
finite $H$. As predicted, while the width of the $\phi^2$
distribution depends strongly on the noise amplitude, the
oscillation amplitude is almost noise independent.

\subsection{The Coupled oscillators model and  NB instability}

In order to imitate the behavior of the NB system with its unstable
fixed point,  Eqs. (\ref{oscillators}) should be modified to include
this new ingredient, manifested here by the terms proportional to
$\kappa$,
\begin{eqnarray}\label{oscillatorsNB}
\frac{\partial x_1}{\partial t} &=& \kappa x_1 + \omega(x_1,y_1) y_1
+ D_1 (x_2-x_1) + \eta_1(t) \nonumber
\\ \nonumber \frac{\partial x_2}{\partial t} &=&  \kappa x_2 +
\omega(x_2,y_2) y_2+ D_1 (x_1-x_2)+ \eta_2(t)  \\ \frac{\partial
y_1}{\partial t} &=&
\kappa y_1 -\omega(x_1,y_1) x_1 + D_2 (y_2 - y_1)+ \eta_3(t) \\
\nonumber \nonumber \frac{\partial y_2}{\partial t} &=& \kappa y_2
-\omega(x_2,y_2) x_2 + D_2 (y_1 - y_2) + \eta_4(t).
\end{eqnarray}
The system is still invariant with respect to global rotation, thus
it reduces to the three dimensional phase space:
\begin{eqnarray}\label{3dNB}
 \dot{R} &=& \left( \kappa -2D sin^2(\phi/2) \right) R + \tilde{\eta}_R  \\
 \dot{r} &=& \left( \kappa -2Dcos^2(\phi/2) \right) r + \tilde{\eta}_r  \\
 \dot{\phi} = -2&D&
 \left( \frac{R^2+r^2}{R^2-r^2} \right) sin \phi +
\omega(r_2) - \omega(r_1) +  \left( \frac{\tilde{\eta}_{1}}{r_1} -
\frac{\tilde{\eta}_{2}}{r_2} \right). \label{8}
\end{eqnarray}
Close to the invariant manifold, thus,
\begin{equation}
P(r) \sim exp[-(2D-\kappa))r^2/\Delta^2]
\end{equation}
and
\begin{equation}
\langle \phi^2 \rangle \sim \frac{\omega^{'}(r)^2 \Delta^2 }{
D(2D-\kappa)^2}.
\end{equation}
Consequently:
\begin{equation}
\langle R^2 \rangle \sim \frac{\Delta^2 }{ D \langle \phi^2 \rangle
- \kappa } = \frac{\Delta^2 }{ \omega^{'}(r)^2
\Delta^2/(2D-\kappa)^2  - \kappa}.
\end{equation}

\begin{figure}
 \includegraphics[width=8cm]{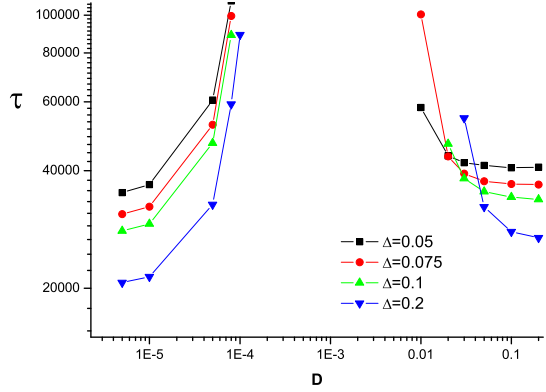}
 \caption{Extinction times for the unstable,   coupled oscillator
 cartoon of the Nicholson-Bailey model. Eqs. (\ref{oscillatorsNB})
 were integrated numerically (using Euler integration with $\Delta
 t = 0.002$) for $\kappa = 0.0001$ and and $\omega = \omega_0 +
 r/2$ for different noise amplitudes $\Delta$ and diffusion constant D.
 The extinction time is plotted for four different noise level
 against the diffusion constant, and the two transitions are
 implicit. The lifetime of the system for large noise
 ($\Delta =0.1, 0.2$, triangles) diverges beyond our computational
 abilities for $D=0.01$.
   }
 \label{nbd}
\end{figure}

\begin{figure}
 \includegraphics[width=8cm]{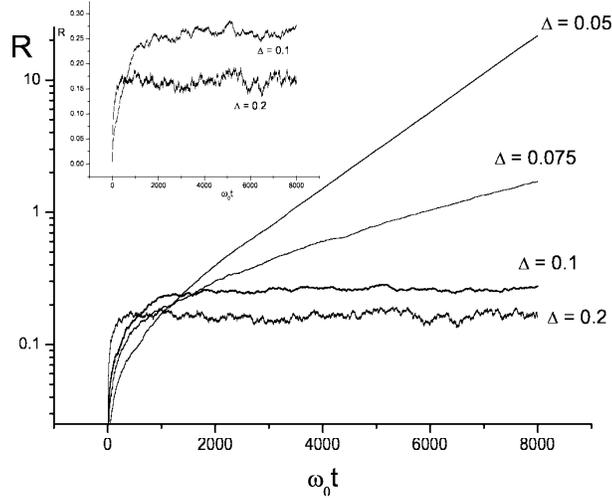}
 \caption{Noise-induced transition for the coupled oscillator
 cartoon of the Nicholson-Bailey model. Eqs. (\ref{oscillatorsNB})
 were integrated numerically (using Euler integration with $\Delta
 t = 0.002$) with $\kappa = 0.0001$, $D = 0.01$ and $\omega = \omega_0 +
 r/2$ for different noise amplitudes $\Delta$. The total distance
 $R$ (averaged over 100 runs) is presented, in logarithmic scale, against
time measured in
 units of $\omega_0$. Small noises are followed by exponential
 growth of the oscillation amplitude, as suggested by the
 deterministic part of (\ref{oscillatorsNB}). The larger the noise,
 the slope of this diverging line becomes smaller. If the noise is
 large enough, $R$ saturates at a finite value, as seen more clearly
 when the scale is not logarithmic (inset).
   }
 \label{nbco1}
\end{figure}

The system becomes unstable when either $r^2$ or $R^2$ diverge. The
first criterion for stability comes from the amplitude
synchronization parameter, $2D > \kappa$, so the diffusion should
increase above some threshold value in order to prevent
desynchronized extinction where the system acts as if made of two
disconnected patches. As in the LV case, if the migration rate is
too large (i.e., if $ \kappa$ becomes larger than $
[\omega^{'}(r)]^2 \Delta^2/(2D-\kappa)^2 \sim \omega'^2 \ \Delta^2
/D^2  $), the system synchronizes and the deterministic flow leads
to synchronized extinction.  Note that close to the low D transition
the extinction rate grows with the noise, while close to the second
transition, increase of noise amplitude $\Delta$ yields lower
extinction rates, emphasizing the fact that the stability is
\emph{noise-induced}. This feature is clearly seen in Figure
\ref{nbco1}, where for small noise the oscillation amplitude grows
exponentially in time while for large noise the oscillation
stabilizes.

Along this section, the various numerical  implementation of the toy
assumes, for simplicity, that the oscillation frequency depends
linearly on the slope.  If, on the other hand, $\omega'$ is a
nonuniform function of the amplitude, oscillations will  grow
sublinearly  (for the LV model) or exponentially (in NB case) with
time, until they reach a phase space region where $\partial \omega /
\partial r$ is large enough. For an LV-like system it leads
to the appearance of a "soft" limit cycle - exponential convergence
from the outside, diffusion inside.

It should be noted that the stability mechanism presented here is
applicable not only for the classic predator-prey systems but for
any biological system that supports, locally, neutrally stable or
unstable oscillations.  In particular it may stabilize a "locally"
unstable ecology of interspecific competition for common resource.
Such a system may be  described mathematically by the  generalized
Lotka-Volterra equations (Chesson, 2000); even if the theory
predicts extinction, species diversity may be maintained on spatial
domains. Another possible application is the stabilization of
"eigenperiods"
 - oscillations due to the interaction between
an individual and its internal resources or by maternal effect -
suggested recently as a predation free mechanism for population
oscillations (Ginzburg  and Colyvan, 2004).

\section{Other Mechanisms}

This section is devoted to a discussion of previously suggested
stabilization mechanisms in the framework of the coupled oscillator
toy model. We show that all the significant mechanisms surveyed by
(Briggs and Hoopes , 2004) become very clear and allow  analytical
understanding when presented using the coupled oscillator system.

\subsection{Spatial heterogeneity}

To implement the spatial heterogeneity (SH) (Murdoch and Oaten ,
1975; Murdoch et al, 1992; Hassell and May, 1988) in our framework,
consider a two patch system with different (radius independent)
frequencies, $\omega_1$ and $\omega_2$:
\begin{eqnarray}\label{ih}
\frac{\partial x_1}{\partial t} &=& \omega_1 y_1 + D_1 (x_2-x_1) +
\eta_1(t) \nonumber
\\ \nonumber \frac{\partial x_2}{\partial t} &=& \omega_2 y_2+
D_1 (x_1-x_2)+ \eta_2(t)  \\ \frac{\partial y_1}{\partial t} &=&
-\omega_1 x_1 + D_2 (y_2 - y_1)+ \eta_3(t) \\ \nonumber \nonumber
\frac{\partial y_2}{\partial t} &=& -\omega_2 x_2 + D_2 (y_1 - y_2)
+ \eta_4(t).
\end{eqnarray}
As these equations are linear, one may again diagonalize the
deterministic part of the evolution matrix to obtain the exact
probability distribution function.  However, as this problem is
still invariant under global rotation and is independent of
$\theta_1+\theta_2$ we have chosen to use the more intuitive polar
representation. After the standard coordinates transformation, one
finds:

\begin{eqnarray}\label{3d3}
 \dot{R} &=& -2D sin^2 \left( \frac{\phi}{2} \right) R + \tilde{\eta}_R  \\
 \dot{r} &=& -2Dcos^2\left( \frac{\phi}{2} \right)r + \tilde{\eta}_r  \\
 \dot{\phi} = -2&D&
 \left( \frac{R^2+r^2}{R^2-r^2} \right) sin \phi +
[\omega_2 - \omega_1] +  \left( \frac{\tilde{\eta}_{1}}{r_1} -
\frac{\tilde{\eta}_{2}}{r_2} \right). \label{73}
\end{eqnarray}

Here $\Delta \omega \equiv \omega_2 - \omega_1$ is a constant, so
close to the invariant manifold, $r$ decouples from $\phi$.
Consequently, while $P(r)$ is still given by \ref{cc}, $\phi$
satisfies an equation for  a \emph{forced} overdamped pendulum and
$sin(\phi) = \Delta \omega/2D$. The "spring constant" in the
invariant manifold $R$ is thus \emph{noise-independent} and
\begin{equation}
\langle R^2 \rangle \sim \frac{D \Delta^2}{(\Delta \omega)^2}.
\end{equation}
The radius of oscillation in a heterogenous system increases with
noise: the system  becomes more extinction prone as the noise grows
larger. In the corresponding "Nicholson-Bailey" toy model the phase
transition happens at $\kappa = D/(\Delta \omega)^2$, and the
location of that transition is \emph{noise-independent}.

\subsection{Environmental stochasticity}

The above framework may also  be used to consider the stabilizing
effect of spatio-temporal environmental stochastisity (STES)
(Crowley, 1981; Reeve, 1990; Taylor, 1998). If on both patches the
radial velocity takes the form $\omega +\zeta_i(t)$, where $i$ is
the patch index and $\zeta$ is a white noise that satisfies $\langle
\zeta(t) \zeta(t') \rangle = \Upsilon \delta(t-t')$, the $\phi$
variable obeys an equation for an overdamped noisy pendulum and the
resulting motion on the invariant manifold satisfies:
\begin{equation}
\dot{R} = \left( \kappa - \Upsilon \right) R + \eta_R,
\end{equation}
so $\langle R^2 \rangle \sim \Delta^2/(\Upsilon - \kappa)$. In that
case the lifetime of the system is \emph{diffusion-independent},
growing with the environmental stochasticity and decaying with the
noise.

\subsection{Jansen instability}

About ten years ago, Jansen (Jansen, 1995; Jansen and de Roos, 2000;
Jansen and Sigmund, 1998) put forward the idea of linearly unstable
orbits of the Lotka-Volterra dynamics, i.e., orbits in the
homogenous manifold for which the highest absolute value of an
eigenvalue of the Floquet operator is larger than 1. This may happen
only if the migration properties of the prey and the predator are
different. In the case of equal diffusivities, the migration term
factors out from the Floquet operator and the stability properties
of orbits lying in the invariant manifold are the same as their
matching trajectories on a single patch (Abarbanel, 1995). However,
Jansen pointed out that if one sets $D_b =0$ in Eqs. \ref{two}, some
orbits may become unstable. In that case one may use the fact that
the total $H$
\begin{equation}\label{Htotal}\
H_{T} \equiv H_1 + H_2  = a_1 + a_2 + b_1 + b_2 - ln(a_1 a_2 b_1
b_2),
\end{equation}
with the deterministic dynamics (\ref{two}) is a \emph{monotonously
decreasing} quantity in the non negative population regime:
\begin{equation} \label{Hdynamics}
\frac{d H_T }{dt} = - D_a \left( \frac{(a_1 - a_2)^2}{a_1 a_2}
\right) - D_b \left( \frac{(b_1 - b_2)^2}{b_1 b_2} \right) < 0.
\end{equation}
Accordingly, if an orbit on the invariant manifold becomes unstable,
the flow will be inward and the population oscillations stabilize.

\begin{figure}
 \includegraphics [width=8cm] {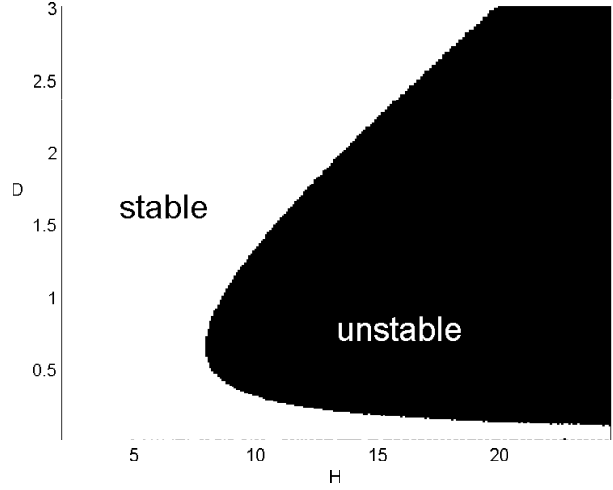}\\
 \caption{Stability diagram for phase space orbits (ordered by their
conserved quantity $H$)
 for different values of predator diffusion $D_a$, where Db=0.}
 \label{jans}
\end{figure}

With the transformation,
\begin{eqnarray}\label{transform}
A=\frac{a_1 + a_2}{2} \qquad B=\frac{b_1 + b_2}{2}\\\nonumber
\delta=\frac{a_1 - a_2}{2} \qquad \theta=\frac{b_1 - b_2}{2},
\end{eqnarray}
one realizes the homogenous $AB$ manifold and the $\delta-\theta$
coordinates that measures the deviation from that manifold (the
heterogeneity of the population). In these coordinates the system
satisfies,
\begin{eqnarray}\label{lv2trans}
\frac{\partial A}{\partial t} &=& - \mu A + \lambda_1  A B +
\lambda_1
\delta \theta\\
\nonumber \frac{\partial B}{\partial t} &=& \sigma B - \lambda_2
A  B + \lambda_2  \delta \theta \\
\nonumber \frac{\partial \delta}{\partial t} &=& - \mu \delta +
\lambda_1 A \theta + \lambda_1 B \delta -2 D_a \delta \\ \nonumber
\frac{\partial \theta}{\partial t} &=& \sigma B - \lambda_2  A
\theta - \lambda_2  B \delta -2 D_b \theta.
\end{eqnarray}
Linearizing around the homogenous manifold, the $AB$ dynamic is
equivalent to that of a  single patch,
\begin{eqnarray}\label{lv2translin}
\dot {A} &=& - \mu A + \lambda_1  A B\\
\nonumber \dot{B} &=& \sigma B - \lambda_2 A  B  \\ \nonumber
\end{eqnarray}
and the $\delta-\theta$ linearized dynamic is
\begin{eqnarray}\label{matrix}
 \frac{\partial}{\partial t} \left(
 \begin{array}{cc}
   \delta \\ \theta \\
 \end{array}
\right) =\left(
 \begin{array}{cc}
   -\mu+\lambda_2 A -2D_a & \lambda_2 \\
   -\lambda_2 A & \sigma - \lambda_2 B -2D_b \\
 \end{array}
\right) \left(
 \begin{array}{cc}
   \delta \\ \theta \\
 \end{array}
\right).
\end{eqnarray}
One may thus calculate the eigenvalues of the Floquet operator for
one period along an orbit of (\ref{lv2translin}). The resulting
stability diagram, first obtained by (Jansen, 1995), is shown in
Figure \ref{jans}.

We have discussed Jansen's stabilization mechanism in a different
publication (Abta and Shnerb, 2006b). It turns out that the
underlying mechanism has to do with the dependence of the
\emph{angular velocity} along the orbit on the azimuthal angle (see
Figure \ref{ang}),  and we can mimic that behavior using our toy
model with $\omega(\theta)$. Specifically, the coupled oscillator
model with,
\begin{eqnarray}\label{co2}
 \frac{\partial x_1}{\partial t} &=&  \omega (\theta_1) y_1 + D_x
(x_2-x_1)\\
 \nonumber \frac{\partial x_2}{\partial t} &=&  \omega (\theta_2) y_2 + D_x
(x_1-x_2) \\
\nonumber \frac{\partial y_1}{\partial t} &=&  -\omega (\theta_1)
x_1 +
D_y(y_2-y_1) \\
 \nonumber \frac{\partial y_2}{\partial t} &=&  - \omega (\theta_2) x_2 +
D_y (y_1
 -y_2),
 \end{eqnarray}
where $D_x = D, \  D_y = 0$ and
\begin{eqnarray}\label{omegadef}
 \omega &=& \omega_0 + \omega_1 \cos(\theta-\frac{\pi}{4}),
 \end{eqnarray}
leads to the same type of instability.

To prove that this model actually yields Jansen's instability, we
have used ($i \in 1,2$)
\begin{eqnarray}\label{tran} r_i^2 = x_i^2 + y_i^2 &\qquad&
 \theta_i  = arctan (\frac{y_i}{x_i}) \\
 \dot{r} = \frac {(x \dot{x} + y \dot{y})}{r} &\qquad&
 \dot{\theta} = \frac { (x \dot{y} - y \dot{x})}{r^2}, \nonumber
\end{eqnarray}
and
\begin{eqnarray}\label{transformation}
 r = r_2-r_1  &\qquad&
 R = r_2+r_1\\
 \phi = \theta_2-\theta_1 &\qquad&
 \Phi = \theta_2+\theta_1. \nonumber
\end{eqnarray}
The flow in the invariant manifold satisfies,
\begin{eqnarray}\label{proximity2}
 \dot R &=& 0 \\
 \nonumber \dot\Phi &=& \omega(\theta_1) + \omega(\theta_2), \\  \nonumber
 \end{eqnarray}
while the linearized equations for the desynchronization amplitude
$r$ and the desynchronization angle $\phi$ satisfy:
\begin{eqnarray}\label{mat}
\frac{\partial}{\partial t} \left(
 \begin{array}{cc}
   \phi \\ r \\
 \end{array}
\right)= \left(
 \begin{array}{cc}
   2\omega'(\Phi/2)-4D_x \cos^2(\Phi/2) & \frac{2D_x \sin\Phi}{R} \\
   -D_x R\sin(\Phi) & -4D_x \sin^2(\Phi/2) \\
 \end{array}
\right)\left(
 \begin{array}{cc}
   \phi \\ r \\
 \end{array}
\right).
 \end{eqnarray}

\begin{figure}
 \includegraphics [width=8cm] {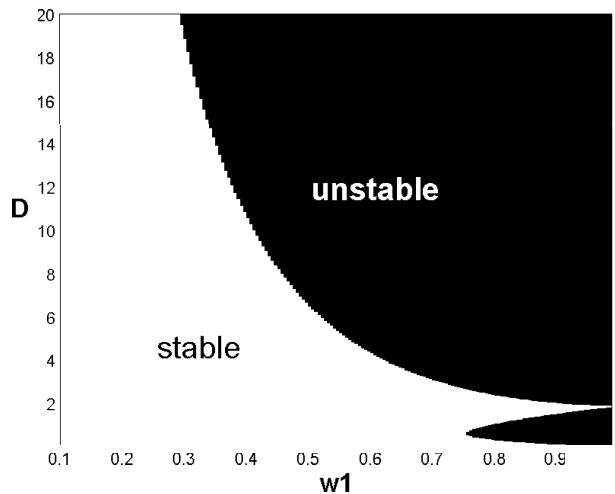}\\
 \caption{Stability diagram in the $\omega_1-D_x$ plane for the Floquet
operator (same  as Figure \ref{jans})
  for the coupled oscillator system described by Eqs. (\ref{co2})
  with $D_y = 0$. The two unstable regions correspond to different
  signs
  of the Floquet unstable eigenvalue, as explained in the text.}
 \label{toy}
\end{figure}

Using the Floquet operator technique  to analyze the stability of an
 orbit by integrating (\ref{mat}) along a close trajectory of
 (\ref{proximity2}), one finds the stability map presented in Figure
 \ref{toy}, where the parameter $H$ of Figure \ref{jans} is now
 replaced by $\omega_1$, which measures the "eccentricity" of the
 angular velocity along a circular path.  Here, two unstable regions
 appear, for large and small $D_x$, but the qualitative picture is
 almost the same.

\section{Concluding remarks: Towards classification of sustained
oscillations.}

Up until now, four different mechanisms that induce stability of
population oscillations for metapopulations have been presented. The
generic mechanism relies on the dependence of the oscillation
frequency on their amplitude (ADF, Amplitude Dependent Frequency).
The other three are spatial heterogeneity (SH), spatio-temporal
environmental stochasticity (STES), and Jansen's mechanism based on
the dependence of the angular velocity on the azimuthal angle.

In natural systems, in the laboratory, and in simulations,  one may
encounter population oscillations in the   presence of  a few of the
above mentioned factors, simultaneously. The task of the researcher
is to make a distinction between them and to identify the relevant,
or the most relevant, stabilizing mechanism. To do that, the
following comments may be useful:

\begin{itemize}

\item Some of the parameters involved, such as the migration rate
per capita, the strength of the environmental stochasticity, growth
and competition parameters and so on, may be known from the
literature or may be measured independently. If possible, an
estimate of the stable oscillation amplitude has to be made and a
comparison between possible mechanisms will reveal  the dominant
one, which is (in the long run) the one that stabilizes the
\emph{smallest} oscillation amplitude.

\item If the dominant mechanism is spatial heterogeneity,  the phase
between cycles on neighboring patches is kept constant in time.
Thus, it is quite easy to recognize an  SH-induced stability. In all
other cases there is no preference among patches and $\phi$ is
distributed randomly around zero.

\item Jansen's RVA instability depends very much on the difference
between species diffusivities. It is impossible to observe such an
instability in host-parasite systems, where the parasite migration
depends on host movement.

\item One way to distinguish between STES and ADF
is to separately  manipulate   the additive   noise (taking small
portions of the population in and out of the system at random) and
environmental stochasticity. An increase  in the amplitude $\Delta$
of the  additive noise leads to larger oscillations for STES and
smaller/equal  oscillation amplitudes for ADF; amplified
stochasticity will diminish oscillations if the basic mechanism is
STES, and is neutral in ADF-controlled systems. Another test is the
response of the system to changes in the migration rate, where STES
synchronized oscillation amplitudes are not affected by changes in
$D$.

\item In both SH and STES mechanisms, the amplitude desynchronization
\emph{decouples} from the phase desynchronization, while in ADF they
are intimately connected. Measurements of the correlations between
the phase and the amplitude desynchronization will  immediately
reveal  the relevant mechanism.
\end{itemize}

As pointed out by Jansen and Sigmund (Jansen and Sigmund, 1998),
"all models of ecological communities are approximations: it is
pointless to burden them with too many contingencies and details. On
the other hand, they would be of little help if they were not robust
against the kind of perturbation and shocks to which a real
ecosystem is ceaselessly exposed." The current accuracy of data on
population oscillations and inter-specific interaction, both from
field studies and from experiments, is, as far as we know, far
beyond the level needed for an exact comparison with theoretical
predictions about the oscillation phase portrait, like those
predicted by Lotka-Volterra and other models. Given that, the main
insights from the available data are, first, the mere existence of
these oscillations, and second, the identification of the underlying
mechanism that limits the amplitude of these oscillations in noisy
environments.  As emphasized by the recent experiments of (Kerr et
al., 2002 ; Kerr et al., 2006) , one may observe persistent
oscillations or extinction, but it is hard to compare the exact
population dynamic with the predictions of the theory. Accordingly,
the analysis of data on population cycles may be preformed, as we
have shown here, completely within the framework of the simple
coupled oscillation model that allows all the suggested limiting
processes within a transparent and general modeling scheme.

\section{Acknowledgements}

 We acknowledge helpful discussions with David Kessler, Uwe
T$\ddot{\textrm{a}}$uber, and  Arkady Pikovsky. This work was
supported by the Israeli Science Foundation (grant no. 281/03) and
the EU 6th framework CO3 pathfinder.

\newpage
\begin{flushleft}
{\bf REFERENCES}

\begin{list}{}{}

\item Abarbanel H.D.I., 1995. \emph{Analysis of observed chaotic data}
Springer, Berlin, p. 87.

\item Abta R., Schiffer M. and Shnerb M.N., 2006a, Amplitude dependent frequency,
desynchronization, and stabilization in noisy metapopulation.
dynamics. cond-mat/0608108.

\item Abta R. and Shnerb M.N., 2006b. Angular velocity variations and stability
of spatially explicit prey-predator systems. q-bio.PE/0612021.

\item  Allen, J.C., 1975. Mathematical models of species interactions in
space and time. \emph{Am. Nat.} \textbf{109}, 319–342.

\item  Bettelheim E., Agam O. \&  Shnerb N.M., 2000.  "Quantum phase
transitions"" in classical nonequilibrium processes, \emph{Physica}
\textbf{E 9}, 600-608.

\item  Blasius B., Huppert A. \&  Stone L., 1999. Complex dynamics and
phase synchronization in spatially extended ecological systems.
\emph{Nature} \textbf{399}, 354-359.

\item  Briggs C.J. \&  Hoopes M.F., 2004. Stabilizing effects in spatial
parasitoid–host and predator–prey models: a review.
 \emph{Theoretical Population Biology} \textbf{65}, 299-315.

\item Chesson P., 2000. Mechanisms of maintenance of spatial
diversity. Annu. Rev. Ecol. Syst. 31, 343-366.

\item  Crowley, P.H., 1981. Dispersal and the stability of predator– prey
interactions. Am. Nat. 118, 673–701.

\item  Cuddington K., 2001. The "balance of nature" metaphor and
equilibrium in population ecology. \emph{Biology and Philosophy}
\textbf{16}, 463-479.

\item  Earn D.J.D., Levin S.A. \& Rohani P., 2000. Coherence and
Conservation. \emph{Science} \textbf{290}, 1360-1363.

\item  Elton C.S., 1924. Periodic Fluctuations in the Numbers of Animals: Their Causes and Effects.
 Br. Jour. Exp. Biol. \textbf{2}; 119-163.

\item  Foley P., 1997. \emph{extinction models for local population}, in
\emph{Metapopulation Biology - Ecology, Genetics ana Evolution},
I.A. Hanski and M.E. Gilpin (Ed.) Academic Press (London).

\item  Freeman D.B., 1992. Prickly pear menace in eastern Australia
1880-1940. Geographical Review \textbf{82}, 413-429.

\item  Gardiner C.W., 2004. \emph{Handbook of Stochastic Methods},
Springer, Berlin.

\item  Gause G.F., 1934. \emph{The struggle for existence}. William  and
Wilkins, Baltimore.

\item  Gillespie  D.T., 1977. Exact stochastic simulation of coupled
chemical reactions . \emph{Journal of Physical Chemistry},
\textbf{81}, 2340-2361.

\item  Ginzburg L. and Colyvan M., 2004.  \emph{Ecological Orbits}, Oxford
University Press, Oxford U.K.

\item  Grassberger. P., 1982. On phase transitions in Schlogl's model.  Z.
Phys. B: Condens. Matter \textbf{47}, 365-374.

\item  Grimm V. and Wissel C., 2004. The intrinsic mean time to extinction:
a unifying approach to analysing persistence and viability of
populations. Oikos \textbf{105},  501-511.

\item  Hassell, M.P., May R.M., 1988. Spatial heterogeneity and the
dynamics of parasitoid–host systems. \emph{Ann. Zool. Fenn.}
\textbf{25}, 55–62.

\item  Hassell, M.P., Varley, G.C., 1969. New inductive population model
for insect parasites and its bearing on biological control.
\emph{Nature} 223, 1133–1137.

\item  Holyoak M. \& Lawler S.P., 1996. Persistence of an Extinction-Prone
Predator-Prey Interaction Through Metapopulation Dynamics.
\emph{Ecology} \textbf{77}, 1867-1879.

\item Huffaker C.B., 1958. Experimental studies on predation: dispersion
factors and predator prey oscillations. Hilgardia \textbf{27}
343-383.

\item Jansen, V.A.A., 1995. Regulation of predator– prey systems through
spatial interactions: a possible solution to the paradox of
enrichment.  \emph{Oikos} \textbf{74}, 384-390.

\item Jansen V.A.A and de Roos A.M., 2000. In: \emph{The Geometry of
Ecological Interactions: Simplifying Spatial Complexity,} eds.
Dieckmann U., Law R and Metz J.A.J., pp. 183 Cambridge University
Press.

\item Jansen V.A.A. and Sigmund K., 1998. Shaken Not Stirred: On
Permanence in Ecological Communities. Theo. Pop. Bio. \textbf{54},
195-201.

\item Janssen H.K., 1981. On the nonequilibrium phase transition in reaction-diffusion systems with an absorbing stationary
state.
 Z. Physik. \textbf{42}, 151-154.

\item  Kerr B., Neuhauser C.,  Bohannan B.J.M. \& Dean A.M., 2006. Local migration promotes competitive
restraint in a host-pathogen 'tragedy of the commons'. \emph{Nature}
\textbf{442}, 75-78.

\item Kerr B.,  Riley M.A.,  Feldman M.W. and  Bohannan B.J.M., 2002.
Local dispersal promotes biodiversity in a real-life game of
rock-paper-scissors. Nature \textbf{418}, 171-174.

\item Kessler D.A.  and Shnerb M.N., 2006. Extinction Rates for Fluctuation-Induced
Metastabilities: A Real-Space WKB Approach. q-bio.PE/0611049.

\item Kirkup  B.C. and  Riley M.A., 2004. Antibiotic-mediated antagonism
leads to a bacterial game of rock-paper-scissors in vivo.
\emph{Nature} \textbf{428} 412-414

\item Kolmogorov A., 1936. Sulla teoria di Volterra della lotta per
l'esistenze. G. Ins. Ital. Attnari \textbf{7} 985-992.

\item Lande R., Engen S. and Saether B.E., 2003. \emph{Stochastic
Population Dynamics in Ecology and Conservation}, Oxford University
Press, Oxford U.K.

\item Lotka A.J. 1920. Analytical note on certain rhythmic relations in
organic systems. \emph{Proc. Natl. Acad. Sci. USA} \textbf{6},
410-415.

\item Luckinbill L.S., 1974. The Effects of Space and Enrichment on a
Predator-Prey System. Ecology, 1142-1147.

\item May R.M., 1972. Limit cycles in predator prey communities. Science
\textbf{177} 900-902.

\item May, R.M.,1978. Host-parasitoid systems in patchy environments: a
phenomenological model. J. Anim. Ecol. \textbf{47}, 833-843.

\item Mobilia M., Georgiev I.T. and Tauber U.C., 2006. Fluctuations and
Correlations in Lattice Models for Predator-Prey Interaction. Phys.
Rev. \textbf{E 73},  040903.

\item Murdoch, W.W., Briggs, C.J., Nisbet, R.M., Gurney, W.S.C., Stewart-
Oaten, A., 1992. Aggregation and stability in metapopulation models.
\emph{Am. Nat.} \textbf{140}, 41–58.

\item  Murdoch, W.W., Oaten, A., 1975. Predation and population
stability. \emph{Adv. Ecol. Res.}\textbf{ 9}, 1–131.

\item Murray J.D.,1993 \emph{Mathematical Biology}. (Springer-Verlag,
New-York).

\item Nicholson, A.J.,  1933. The balance of animal populations. \emph{ J.
Anim. Ecol. }\textbf{2}, 132-178.

\item Nicholson, A.J., Bailey, V.A. 1935. The balance of animal
populations \emph{Proc. Zool. Soc. London} Part I 3, 551-598.

\item Pimentel D., Nagel W.P. \& Madden J.L., 1963. Space-time structure
of the environment and the survival of parasite-host systems
American Naturalist, \textbf{97}, 141-167.

\item Reeve, J.D., 1990. Stability, variability, and persistence in host–
parasitoid systems. \emph{Ecology} \textbf{71}, 422–426.

\item Redner S., 2001. \emph{A Guide to First-Passage Processes},
Cambridge University Press, Cambridge.

\item Reichenbach T., Mobilia M.  and  Frey E., 2006. Coexistence versus
extinction in the stochastic cyclic Lotka-Volterra model. Phys. Rev.
E \textbf{74}, 051907.

\item Rosenzweig, M.L., MacArthur, R.H., 1963. Graphical representation
and stability conditions of predator–prey interactions. Am. Nat. 97,
209–223.

\item Stenseth et. al., 1998. From patterns to processes: Phase and
density dependencies in the Canadian lynx cycle. Proc. Natl. Acad.
Sci. USA \textbf{95}, 15430-15435.

\item Taylor A.D., 1990. Metapopulations, Dispersal, and Predator-Prey
Dynamics: An Overview. Ecology \textbf{71}, 429-433.

\item Taylor A.D., 1998. Environmental variability and the persistence of
parasitoid–host metapopulation models. Theo. Pop. Biol. 53, 98-107.

\item Volterra V, 1931. \emph{Lecon sur la Theorie Mathematique de la
Lutte pour le via, Gauthier-Villars}, Paris.

\item Washenberger M.J.,  Mobilia M. \& T$\rm{\ddot{a}}$uber U.C., 2006.
Influence of local carrying capacity restrictions on stochastic predator-prey models.
 cond-mat/0606809.

\item Wilmers C.C, Post E., Peterson R.O. \& Vucetich J.A, 2006. Predator
disease out-break modulates top-down, bottom-up and climatic effects
on herbivore population dynamics. Ecology Letters, \textbf{9}
383–389.

\item Wilson, W.G., de Roos, A.M., McCauley, E., 1993. Spatial instabilities
within the diffusive Lotka–Volterra system: individual-based
simulation results. Theor. Popul. Biol. 43, 91–127.

\end{list}

\end{flushleft}

\end{document}